\documentclass{article}

\usepackage[final]{neurips_2021}

\usepackage{geometry}
\usepackage{longtable}
\usepackage{array}
\usepackage[table,xcdraw]{xcolor}
\usepackage[table,xcdraw]{xcolor}
\usepackage[utf8]{inputenc} 
\usepackage[T1]{fontenc}    
\usepackage{hyperref}       
\usepackage{url}            
\usepackage{booktabs}       
\usepackage{amsfonts}       
\usepackage{nicefrac}       
\usepackage{microtype}      
\usepackage{xcolor}         
\usepackage{graphicx}       

\title{What we learned while automating bias detection in AI hiring systems for compliance with \\ NYC Local Law 144}

\author{%
   Gemma Galdon Clavell  \\
   Founder \& CEO \\
   Eticas.ai \\
   \texttt{gemma@eticas.ai} \\
   \And
  Rubén González Sendino \\
  Lead Data Scientist\\
  Eticas.ai\\
  \texttt{ruben.gonzalez@eticas.ai} \\
}

\begin{document}

\maketitle

\begin{abstract}
    Since July 5, 2023, New York City’s Local Law 144 requires employers to conduct independent bias audits for any automated employment decision tools (AEDTs) used in hiring processes. The law outlines a minimum set of bias tests that AI developers and implementers must perform to ensure compliance. Over the past few months, we have collected and analyzed audits conducted under this law, identified best practices, and developed a software tool to streamline employer compliance. Our tool, ITACA\_144, tailors our broader bias auditing framework to meet the specific requirements of Local Law 144. While automating these legal mandates, we identified several critical challenges that merit attention to ensure AI bias regulations and audit methodologies are both effective and practical. This document presents the insights gained from automating compliance with NYC Local Law 144. It aims to support other cities and states in crafting similar legislation while addressing the limitations of the NYC framework. The discussion focuses on key areas including data requirements, demographic inclusiveness, impact ratios, effective bias,  metrics, and data reliability.
\end{abstract}

\section{Introduction}

At Eticas.ai we have been conducting bias audits and other assessments of AI systems for years. Recently, we have begun to automate our solutions around a software platform, ITACA\_OS, which offers bias measurement, benchmarking and analytics for AI systems deployed in any field. 
 
With the passing of NYC Local Law 144 in 2021, which prohibits ``employers and employment agencies from using an automated employment decision tool unless the tool has been subject to a bias audit within one year of the use of the tool''s, we developed a software platform to facilitate bias measurement following the specifications of the law. Our Local Law 144 compliance solution is a ``lite'' version of our main bias auditing SaaS, ITACA\_OS, called ITACA\_144\footnote{\url{https://eticas.ai/nyc-law-144/}}.  
 
Our use of software and AI to automate compliance has allowed us to achieve two major breakthroughs: 

\begin{itemize}
    \item To facilitate and encourage bias auditing by offering a compliance solution at a fraction of the cost of manual auditing services.
    \item To turn a compliance effort and cost into a product optimization solution that can identify and minimize error rates and boost system accuracy. 
\end{itemize}

We designed our ITACA\_OS solution as a comprehensive, sector-agnostic bias measurement and auditing platform that allows our clients to identify bias in direct and indirect identifiers, measure drift and error rates for several established attributes throughout the AI system, and monitor outlier performance both in training, production and impact data. ITACA\_OS provides a robust bias measurement and identification tool that allows technical and compliance teams to access insights into model performance and build bias prevention in model production.

\section{Learnings \& Discussion}

\subsection{Data requirements }

The current text for Local Law 144 does not specify data requirements. Companies auditing their systems could use any dataset, from anywhere, and for any time range. We have observed some population breakdowns in published audits that lead us to believe that the data being used may not be from NYC, for instance.

We recommend that future iterations of the law include specific data requirements that specify, at least, that: 

\begin{itemize}
\item The dataset used for auditing should only include historical data from the last 12 months alone. 
\item The dataset used for auditing should include data from NYC-relevant hiring processes 
\end{itemize}

These specifications would reduce deployment and temporal bias and ensure that all organizations use comparable data that can be measured against relevant benchmarks and metrics. 
 
\subsection{Demographic inclusiveness}

In its current formulation, the Law systematically excludes American Indian, Alaska Native, Native Hawaiian, Pacific Islander, Two or more race, and Some Other Races communities, and intersectional groups.  

The law states that: ``An independent auditor may exclude a category that represents less than 2\% of the data being used for the bias audit from the required calculations for impact ratio.'' Based on this provision, auditors are excluding categories with a representation below 2\% by justifying that such representation is insufficient for meaningful statistical analysis. 

Table \ref{tab:my-table} illustrates how the population is distributed in New York City. As we can see, there are population groups whose representation is below 2\%. How will this be accounted for in the selection process, given that 2\% is the threshold? If these populations are underrepresented and the law permits their exclusion from the analysis, categories that are often the most vulnerable to algorithmic bias may be omitted. This exclusion allows for the omission of the groups it aims to protect, and undermines the goal of the law, which is to safeguard against bias and protect these populations. 

We recommend the removal of the 2\% rule, and to add further clarity to the definition of ``Some Other Race''. 

\begin{table}[h!]
\centering
\resizebox{1\textwidth}{!}{%
\begin{tabular}{llll}
\textbf{Race}                                                                                   & \textbf{Sex}                  & \textbf{Value}              & \textbf{Percentage}         \\
\rowcolor[HTML]{E5E4E4} 
{\color[HTML]{000000} \textbf{American indian and alaska native alone, not hispanic or latino}} & {\color[HTML]{000000} Female} & {\color[HTML]{000000} 8037} & {\color[HTML]{000000} 0.11} \\
\rowcolor[HTML]{E5E4E4} 
{\color[HTML]{000000} \textbf{American indian and alaska native alone, not hispanic or latino}} & {\color[HTML]{000000} Male}   & {\color[HTML]{000000} 7418} & {\color[HTML]{000000} 0.11} \\
\textbf{Asian alone, not hispanic or latino}                                                    & Female                        & 620127                      & 8.16                        \\
\textbf{Asian alone, not hispanic or latino}                                                    & Male                          & 549193                      & 7.44                        \\
\textbf{Black or african american alone, not hispanic or latino}                                & Female                        & 826838                      & 11.02                       \\
\textbf{Black or african american alone, not hispanic or latino}                                & Male                          & 658609                      & 9.16                        \\
\textbf{Hispanic or latino}                                                                     & Female                        & 1068147                     & 14.81                       \\
\textbf{Hispanic or latino}                                                                     & Male                          & 942004                      & 13.47                       \\
\rowcolor[HTML]{E5E4E4} 
\textbf{Native hawaiian and other pacific islander alone, not hispanic or latino}               & Female                        & 1607                        & 0.02                        \\
\rowcolor[HTML]{E5E4E4} 
\textbf{Native hawaiian and other pacific islander alone, not hispanic or latino}               & Male                          & 1190                        & 0.02                        \\
\rowcolor[HTML]{E5E4E4} 
\textbf{Some other race alone, not hispanic or latino}                                          & Female                        & 48117                       & 0.70                        \\
\rowcolor[HTML]{E5E4E4} 
\textbf{Some other race alone, not hispanic or latino}                                          & Male                          & 44708                       & 0.67                        \\
\rowcolor[HTML]{E5E4E4} 
\textbf{Two or more races, not hispanic or latino}                                              & Female                        & 129450                      & 1.83                        \\
\rowcolor[HTML]{E5E4E4} 
\textbf{Two or more races, not hispanic or latino}                                              & Male                          & 105373                      & 1.57                        \\
\textbf{White alone, not hispanic or latino}                                                    & Female                        & 1209422                     & 15.80                       \\
\textbf{White alone, not hispanic or latino}                                                    & Male                          & 1137971                     & 15.09                      
\end{tabular}%
}
\caption{Information extracted from the Census API for New York City data (2020)  
showing populations with a representation \textless{}2\% }
\label{tab:my-table}
\end{table}
 
\subsection{Impact ratio vs. Fairness}

In the last few years, many developers and researchers have suggested different model fairness metrics to identify and measure bias in AI models. Law 144 requires AI hiring system developers and implementors to calculate only one: impact ratio, which focuses on measuring differences in selection rates between protected categories. We understand this choice, because IR is a ratio that can be applied in production environments without requiring true labels. However, this metric alone is both limited and misleading if the overall goal is to assess whether a system is fair or unfair.  
 
Equal Impact Ratios between groups do not necessarily guarantee fairness. Proportional outcomes indicate alignment in selection rates, but they do not confirm unbiased treatment of individuals or that the model accurately evaluates qualifications or characteristics. 

To thoroughly evaluate a model’s behavior, a deeper analysis is essential. This includes examining whether proxy features or differential behavior based on inherent group characteristics exist. For instance, even if race is not explicitly used in a model, attributes correlated with race, such as spoken language, could serve as proxies. Language, in this context, could enable the inference of race or ethnicity, potentially leading to indirect discrimination. Conducting counterfactual analysis is crucial to fully understand the model’s behavior and ensure that it is entirely free from discriminatory practices. 

Given these limitations, the central question lies in the intent of the law. Does it seek to guarantee proportional outcomes regardless of the model's underlying behavior? Or does it aim to ensure equal treatment for all individuals, independent of existing societal disparities? Clarifying this objective is critical to determining whether the Impact Ratio is the most suitable metric for these audits.

\subsection{Effective bias}

We’d also like to highlight how limited model metrics are to assess whether an AI system is fair or not. Collecting only model metrics and not impact outcomes is like analyzing one ingredient to assess whether a final dish is safe and tasty. It can’t be done. In hiring, the AI model and its metrics are only one piece in a long and complex process of decision-making, which typically includes a screening process, assessments or tests, technical interviews, behavioral interviews, and manager interviews. In our work as auditors, we have seen how technical decisions made before the AI model intervenes, and human decisions made after the AI model produces an output, can have significant impacts on the final outcome of an algorithmic process. Model metrics are always blind to the structural discrimination and bias they inherit in the training and input data, and any data curation taking place. They are also blind to whatever comes after: a model may tell us who received a high ranking, but not who was offered a job, or who accepted it. This limited vision means that any assessment of bias and fairness at the model level will be partial and never able to capture effective bias, nor point developers to what best practices to incorporate to improve outcomes. 

In our audits using our full suite, ITACA\_OS,  we capture data on outlier performance for each attribute in pre-processing, in-processing and post-processing, measuring the impact of both technical and human interventions, but also how societal dynamics are inherited by AI models. This allows us to measure bias from different angles and in different moments and scenarios, ensuring compliance but also that developers have insight on where corrections and mitigation efforts should focus.  

Documentation and transparency requirements for AI hiring developers and implementors should span throughout the life-cycle of the AI system. Without addressing potential biases at every stage, the overall fairness of the hiring system may be compromised, regardless of the quality of the AEDT system.  
 
\subsection{Metrics}

We’d finally like to point to the need for AI policies to work on and enforce metrics. Law 144 relies on an established metric, the 80/20 rule, to assess whether a model is performing within an acceptable range. This metric, however, is not enforced. An audit can find that a model is performing outside the 80/20 range, and no action would be taken. Beyond that, there are other metrics that could better steer AI developers towards better practices. In our ITACA\_OS audits, we use benchmarks based on representativity. 

For audits in NYC, we draw in data from the US census to have updated demographic data and show our clients how far or close they are to making representative hires. We also place the weight of our analysis on the improvement of outlier performance (bias) between training/input data and model outputs, as we believe that our clients should at least aim to improve the representativity of their AI systems when their data inherits bias. These are choices we have made to make sure audits are a tool that does not merely measures, but also sets the standard for what acceptable and desirable means. We would like to see policy-makers and regulators engage in such discussions, to guide AI auditors and clear metrics to AI developers and implementors. 

\subsection{Data Reliability}

Finally, we want to highlight a common challenge in AI auditing, inspection, and assessment: the reliance on data provided by auditees. In our experience, organizations seeking AI audits today demonstrate a strong commitment to compliance and accountability. However, to prevent the emergence of bad practices, mechanisms must be in place to make dishonest reporting difficult. One effective approach would be for regulators requiring independent AI audits to commit to conducting full, in-depth evaluations of a small percentage of audited systems. These evaluations would involve executing the system alongside developers and auditors to verify the provided data and assess counterfactual fairness. Such random sampling and oversight, a practice widely used in other sectors, would help ensure that the data submitted for auditing reflects the actual system in operation. This added layer of scrutiny would serve as a deterrent to bad practices and strengthen the credibility of the audit results.

\section{Conclussions}

As an organization committed to furthering of the AI auditing profession and its methods, we welcome Local Law 144 and any efforts made by public and private actors to ensure that AI systems are safe and fair. Local Law 144 is an important step toward the standardization of independent AI audits as a crucial inspection and accountability mechanism for the AI systems that shape the life chances of millions of people in NYC, the US, and globally. It is also a key precedent that will inspire similar regulatory developments in other US states and elsewhere. As independent AI auditors, at Eticas.ai we are encouraged by these developments and a growing recognition of the need for AI auditing solutions and methods. 
 
However, as we have oftentimes seen the best intentions of regulators turn into inadequate requirements for compliance, we hope that sharing our insights can help develop better bias measurement standards. Measuring the wrong things helps no one: it forces AI developers to invest in compliance exercises that do not build more robust systems, it makes policy-makers focus on accountability exercises that don’t really protect users and leaves citizens and society as a whole vulnerable to the negative impacts of AI innovation.  
 
As we continue to develop software solutions, benchmarks and metrics to audit AI systems, we aim to shape a field that is robust, useful and focused on achieving meaningful results in terms of AI safety, fairness and accountability.

\end{document}